\documentclass[twocolumn,superscriptaddress,amsmath,amssymb,floatfix,aps,pre,showpacs]{revtex4}
\usepackage{amsmath}    
\usepackage{graphicx}   
\usepackage{verbatim}   
\usepackage{color}      
\usepackage{subfigure}  
\usepackage{hyperref}   
\usepackage{mathrsfs}
\usepackage{amssymb,amsfonts,amsmath}    
\usepackage{graphicx}   
\usepackage{verbatim}   
\usepackage{color}      
\usepackage{subfigure}  
\usepackage{hyphenat}
\usepackage[normalem]{ulem}
\usepackage{array}

\usepackage{color}      
\usepackage{soul}

\begin{document}

\title{The accuracy of telling time via oscillatory signals} 

\author{Michele Monti}
\email{monti@amolf.nl}
\affiliation{FOM Institute AMOLF, Science Park 104, 1098 XG Amsterdam,
  The Netherlands}
\author{Pieter Rein ten Wolde}
\email{tenwolde@amolf.nl}
\affiliation{FOM Institute AMOLF, Science Park 104, 1098 XG Amsterdam, The Netherlands}

\date{\today}                                           

\begin{abstract}
  Circadian clocks are the central timekeepers of life, allowing cells
  to anticipate changes between day and night. Experiments in recent
  years have revealed that circadian clocks can be highly stable,
  raising the question how reliably they can be read out. Here, we
  combine mathematical modeling with information theory to address the
  question how accurately a cell can infer the time from an ensemble
  of protein oscillations, which are driven by a circadian clock. We
  show that the precision increases with the number of oscillations
  and their amplitude relative to their noise. Our analysis also
  reveals that their exists an optimal phase relation that minimizes
  the error in the estimate of time, which depends on the relative
  noise levels of the protein oscillations. Lastly, our work shows
  that cross-correlations in the noise of the protein oscillations can
  enhance the mutual information, which suggests that cross-regulatory
  interactions between the proteins that read out the clock can be
  beneficial for temporal information transmission.
\end{abstract}


\maketitle

\section*{Introduction}
Among the most fascinating timing devices in biology are circadian
clocks, which are found in organisms ranging from cyanobacteria and
fungi, to plants, insects and animals. Circadian clocks are
biochemical oscillators that allow organisms to coordinate their
behavior with the 24-hour cycle of day and night. Remarkably, these
clocks can maintain stable rhythms for months or even years in the
absence of any daily cue from the environment, such as light/dark or
temperature cycles \cite{Johnson2008}. In multicellular organisms, the
robustness can be explained by intercellular interactions
\cite{Liu:1997uv,Yamaguchi:2003jj}, but it is now known that even
unicellular organisms can have very stable rhythms. An excellent
example is provided by the clock of the bacterium {\it Synechococcus
  elongatus}, which is one of the most studied and best characterized
model systems \cite{Johnson2008}. This clock has a correlation time of
several months \cite{Mihalcescu:2004ch}, even though the clocks of the
different cells in the population do not seem to interact with one
another \cite{Mihalcescu:2004ch}. Clearly, the clock is designed in
such a way that it has become resilient against the intrinsic
stochasticity of the chemical reactions that constitute the clock
\cite{Zwicker2010,Paijmans2015}. The observation that clocks can be very
stable, suggests that they are also read out reliably. Yet, how
cells could do so is a wide open question \cite{Mugler:2010cq}.

In this manuscript we combine information theory with mathematical
modeling to study how accurately cells can infer time from cellular
oscillators. While our analysis is general,  it is inspired by the
circadian clock of
{\it S. elongatus}. The central clock component of {\it S. elongatus} is KaiC,
which forms a hexamer \cite{Kageyama2003}. KaiC has two
phosphorylation sites per monomer, which are phosphorylated and
dephosphorylated in a well-defined temporal order, yielding a
protein-phosphorylation cycle (PPC) with a 24 hour period
\cite{Rust2007,Nishiwaki2007}.  This PPC is coupled to a
transcription-translation cycle (TTC) of KaiC \cite{Kitayama2008},
which is a protein synthesis cycle with a 24 hr rhythm, via the
response regulator RpaA.  KaiC in the phosphorylation phase of the PPC
activates the histidine kinase SasA, which in turn activates RpaA via
phosphorylation
\cite{Takai2006,Taniguchi:2007jx,Taniguchi2010,Gutu2013}. In contrast,
KaiC that is in the dephosphorylation phase of the PPC and bound to
KaiB, activates the phosphatase CikA, which dephosphorylates and
deactivates RpaA \cite{Taniguchi2010,Gutu2013}.  Active,
phosphorylated RpaA drives genome-wide transcriptional rhythms, which
include the expression of the clock components
\cite{Markson2013}. 

\begin{figure}[t]
    \centering
    \includegraphics[scale=0.5]{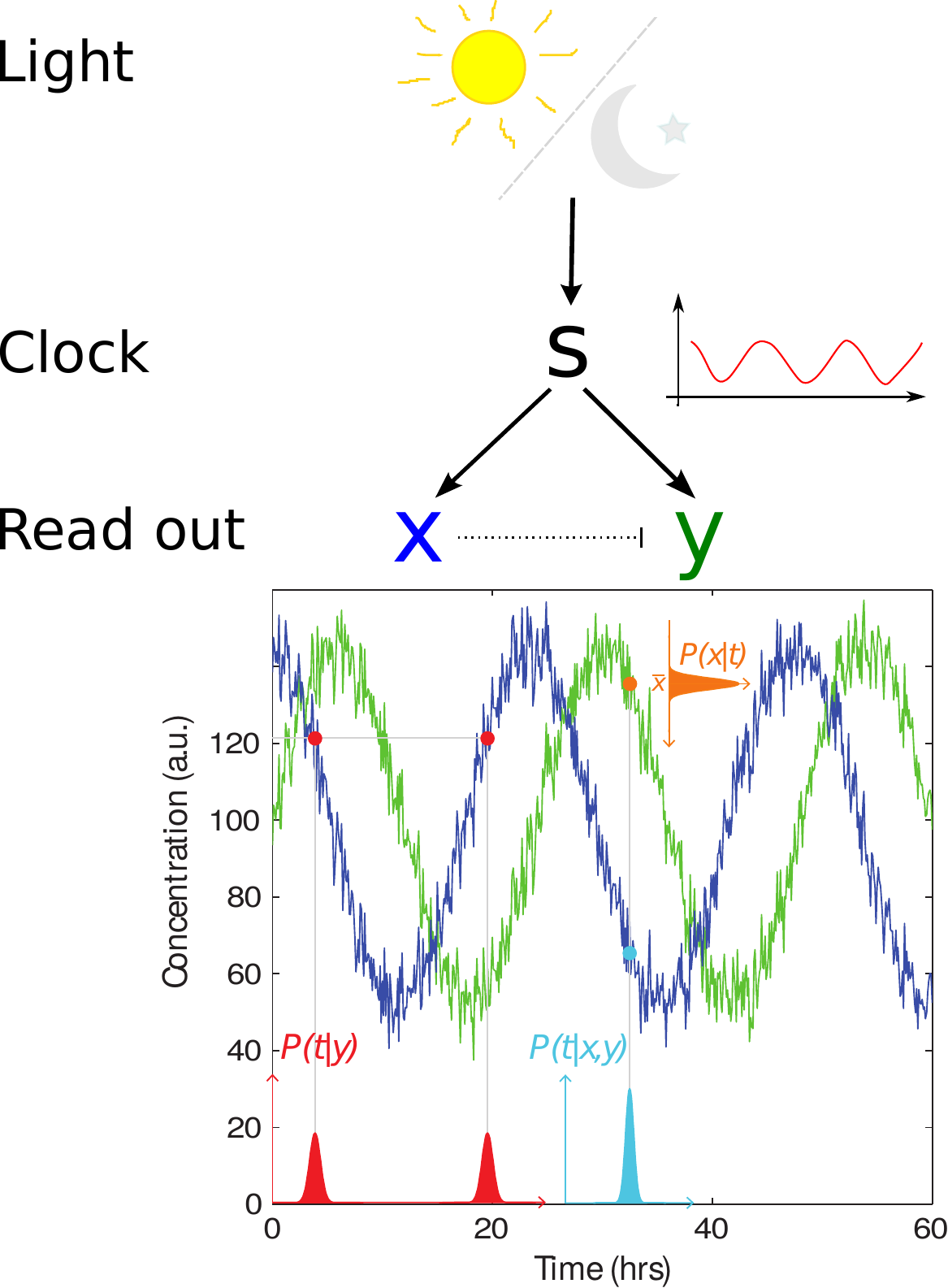}
    \caption{Cells can infer time from an ensemble of protein
      oscillations. The day-night rhythm entrains a circadian clock, a
      biochemical oscillator with a 24 hr period, which in turn drives
      the oscillatory expression of a number of readout genes. The
      precision of the estimate of time depends on the number of
      readout proteins, the amplitude of the
      oscillations, their noise level, their phase difference, and the
      cross-correlations in the fluctuations.}
    \label{fig:fig1}
\end{figure}

Intriguingly, while time could be uniquely encoded in the modification
state of the two phosphorylation sites of KaiC, cells do not seem to
employ this mechanism \cite{Gutu2013,Markson2013}. RpaA, the central
node between the clock and the downstream genes, has only one
phosphorylation site \cite{Gutu2013,Markson2013}. This makes the
question how accurately the cell can infer time a very pertinent one,
because a single readout---the phoshorylation level of RpaA---leads to
an inherent ambiguity in the mapping between time and clock output: a
given level of active RpaA corresponds to two possible times (see
Fig. \ref{fig:fig1}). On the other hand, it is known that RpaA controls the
expression of many downstream genes \cite{Markson2013}. While
their expression levels cannot contain more information about time
than that which is available in the {\it time trace} of RpaA, it is
possible that, collectively, their expression levels do contain more
information about time than that present in the {\it instantaneous} level of
RpaA.

In this manuscript, we study  how the accuracy of telling time depends on the
number of genes that read out a clock, their phase difference, the
level of biochemical noise, and the cross-correlations between the
gene expression levels. In the next section, we first describe the set
up of our analysis, and then the measures that we employ to quantify
information transmision. We then show that there exists an optimal
phase difference that maximizes information
transmission. Interestingly, the optimal phase difference depends on
the amplitude of the noise in the expression of the readout genes, and
on the cross-correlations between them, akin to what has been observed
in neuronal coding \cite{Tkacik2010} and in the gap-gene expression
system of {\it Drosophila} \cite{Walczak:2010cv,Dubuis2013}.

\section{Methods}

\subsection{Model}
\label{sec:Model}
The analysis we present below applies to any readout system that obeys
Gaussian statistics. Yet, to set the stage, and to introduce the key
quantities that we will study, it is instructive to consider a
concrete system. To this end, imagine an oscillatory clock protein,
like RpaA, that drives the expression of a set of downstream genes. Assuming that the system can be linearized, the dynamics of the
system is given by
\begin{equation}
\label{Eq1}
\frac{d{\bf x}(t)}{dt} = {\bf f}s(t) + {\bf B}{\bf x}(t) + \xi (t).
\end{equation}
Here ${\bf x}(t)$ is a vector with components $x_i(t)$, which denote
the concentration $x_i$ of protein ${\rm X}_i$, $s(t)$ is the
concentration of the clock protein, ${\bf f}$ is a vector with
components $f_i$, which describe how the downstream protein ${\rm
  X}_i$ is driven by $s(t)$, ${\bf B}$ is the matrix that describes
the regulatory interactions between the downstream proteins, and
$\xi(t)$ is a vector with components $\xi_i(t)$ that describe the
noise in the expression of ${\rm X}_i$. In what follows, we imagine
that the clock protein osscilates according to $s(t) = A_s \sin(\omega
t) + r_s + \xi_s$, where $A_s$ sets the amplitude of the oscillations,
$r_s$ its mean, and $\xi_s$ describes the noise in the input signal.

This linear system can be solved analytically. For example, if the downstream
genes do not interact with each other and protein ${\rm X}_i$ decays
with rate $\mu_i$, then each protein oscillates as
\begin{equation} \label{xybeh}
	x_i(t) = A_i \sin (\omega t +  \phi_i) + r_i + \eta_i(t),
\end{equation}
where
\begin{eqnarray}
\eta_i(t) &=& \int_{-\infty}^t dt' e^{-\mu_i(t-t')}
\left[\xi_i(t')+f_i \xi_s(t')\right],\label{eq:eta}\\
A_i &=& \frac{f_i A_s}{\sqrt{\mu_i^2 + \omega^2}}, \label{eq:Amp}\\ 
\phi_i &=&\arcsin(\frac{-\omega}{\sqrt{\mu_i^2 + \omega^2}}),\label{eq:phase}\\
r_i &=& \frac{f_ir_s}{\mu_i}.
\end{eqnarray}
  Importantly, even in this
simple system, the difference in the phase $\phi$ between the
expression of the downstream genes can be modulated, namely by
changing the protein degradation rate $\mu_i$. Also the amplitude
$A_i$ can be adjusted; it can be set independently from the phase via
the synthesis rate $f_i$. Both quantities affect the precision by
which the system can estimate the time.

Another key quantity is the noise in the expression of the downstream genes.
Following the linear-noise approximation, we assume that the noise in
the concentration $x_i$ is Gaussian, such that 
\begin{equation}\label{noiseterm}
P(\eta_i) = P(x_i|t^*) = \frac{1}{\sqrt{2 \pi \sigma_{i}^2}}  e^ 
{-\frac{({x_i}-\bar{x}_i(t^*))^2}{2 \sigma_{i}^2}}
\end{equation}
where $\bar{x}_i(t^*)$ is the mean concentration of protein ${\rm
  X}_i$ at time $t^*$, $\sigma^2_{i}=\sigma^2_i(t^*)$ is the variance
of $x_i$ around its mean $\bar{x}_i$, and $t^*$ is the given time. The
noise $\sigma^2_i(t)$ has a extrinsic contribution coming from the
noise in the input signal, an intrinsic contribution from the noise in
the expression of $X_i$, and a contribution from the regulatory
interactions. Our analysis does not depend on the origins of these
noise contributions: in the analysis below, we specify the variance
$\sigma^2_i(t)$ and the co-variance of the fluctuations in $x_i$ and
$x_j$, and then study how this affects the precision of telling time.
In general, we expect, however, that $\sigma^2_i(t)$ depends on the
mean $\bar{x}_i(t)$, and a reasonable assumption is that the variance
equals the mean, $\sigma^2_i(t) = \bar{x}_i (t)$, as in gene
expression via simple Poissonian birth-death statistics \cite{paulsson2003}. However, if the mean $r_i$ of the
protein oscillations is large compared to their amplitude $A_i$, then
we may assume that $\sigma^2_i(t)$ is constant in time, $\sigma^2_i(t)
= \sigma^2_i=r_i$. As will become clear in the next section, the
importance of noise depends on the amplitude of the oscillations: the
key control parameter is the relative noise strength
$\widetilde{\sigma}_i \equiv A_i / \sigma_i$.  This ratio can be
varied independently from gene to gene, $A_i/\sigma_i \neq A_j /
\sigma_j$ in general, and below we will study how this affects the
precision.  If there is no noise in the input $s(t)$ and if the
downstream proteins do not interact with each other (as in the example
considered here), then the cross-correlation between the fluctuations
of the concentrations of the downstream proteins is zero: $\langle
\eta_i \eta_j\rangle = \langle \eta^2_i \rangle
\delta_{ij}=\sigma^2_{i}$, where $\delta_{ij}$ is the Kronecker
delta. However, in general, the noise in the expression of the
downstream genes will be correlated, which, as we will show, can
either enhance or reduce the accuracy by which the downstream proteins
can infer time.

Below, we will consider how the accuracy of telling time depends on
the cross-correlations between the expression of the downstream genes,
their phase difference, and on $\widetilde{\sigma}_i$, and how this varies
from gene to gene. 

\subsection{Reliability measures} 
The central idea of our analysis is that the system infers the time
from the collective expression of the $N$ downstream proteins,
$\{x_i\}\equiv \{x_1(t), x_2(t), \dots, x_{N-1}(t),
x_N(t)\}$. Following work on positional information in {\it
  Drosophila} \cite{Dubuis2013}, we use two approaches to quantify the
accuracy on telling time. The first is based on the error in the
estimate of a given time $t$, $\sigma_t(t)$; a related approach has been widely
used to derive the fundamental limits on the accuracy of sensing
\cite{berg1977,Ueda:2007uq,bialek2005,levinepre2007,levineprl2008,wingreen2009,levineprl2010,mora2010,Govern2012,Mehta2012,Skoge:2011gi,Skoge:2013fq,Kaizu:2014eb,Govern:2014ef,Govern:2014ez,Lang:2014ir}. The
second approach is based on the mutual information, which in recent years has
been used extensively to quantify cellular information transmission
\cite{Ziv2007,Tostevin2009,Mehta2009,Tkacik:2009ta,tostevin10,DeRonde2010,Tkacik2010,Walczak:2010cv,DeRonde2011,Cheong:2011jp,deRonde:2012fs,Dubuis2013,Bowsher:2013jh,Selimkhanov:2014gd,DeRonde:2014fq,Govern:2014ez,Sokolowski:2015km,Becker2015}.

\subsubsection{The error in estimating time}
To determine the error in estimating the time, we start from the
generalization of Eq. \ref{noiseterm} to multiple downstream genes: 
\begin{equation}
P(\{x_i\}|t) = \frac{1}{\sqrt{2 \pi |{\bf  C}|}} \exp\left[-\frac{1}{2}\sum_{i,j}^N \delta x_i  C^{-1}_{ij} \delta x_j\right].
\end{equation}
Here $ \delta x_i (t)= x_i(t) -\bar{x_i} $, ${\bf C}$ is the covariance
matrix with elements $C_{ij}$, $|{\bf C}|$ is its determinant and
${\bf C}^{-1}$ is its inverse.

The idea is now to invert the problem, and ask what is the
distribution of possible times $t$, given that the expression levels
are $\{x_i\}$. This can be obtained from Bayes' rule:
\begin{equation}\label{Ba_ru}
P( t |\{x_i\}) = P(t) \frac{P(\{x_i\}|t)}{P(\{x_i\})}
\end{equation}
where $P(t)=\frac{1}{T}$ is the uniform prior probability of having a
certain time and $P(\{x_i\})$ is the joint distribution of the
expression levels of the downstream genes.  If the noise $\eta$ is
small compared to the mean, then $P(t|\{x_i\})$ will be a Gaussian distribution that is
peaked around $t^*(\{x_i\})$, which is the best estimate of the
time given the expression levels
 \cite{Tkacik2011,Dubuis2013}:
\begin{equation}
P(t|\{x_i\})\simeq  \frac{1}{\sqrt{2 \pi \sigma_t^2}} \exp\left[
-\frac{(t-t^*(\{x_i\}))^2}{2 \sigma_t^2}\right].
\end{equation}
Here $\sigma^2_t=\sigma^2_t(t*)$ is the variance in the estimate of the
time, and it is given by \cite{Dubuis2013}
\begin{equation}\label{eq:sigt}
\sigma_t^{-2} \simeq \sum_{i,j}^N \left.\left[\frac{d\bar{x}_i(t)}{dt} C^{-1}_{ij} \frac{d\bar{x}_j(t)}{dt}\right]\right|_{t=t^*(\{x_k\})}
\end{equation}

We first consider the scenario in which the noise in the expression of
the downstream genes, $\eta_i$, is uncorrelated from one gene to the
next.  In this case ${\bf C}$ is a diagonal matrix where the diagonal
elements are the variances of the respective protein concentrations:
$C_{ii} = \sigma_{i}^2$.  Substituting $C_{ii}$ and Eq \ref{xybeh} in
Eq \ref{eq:sigt} we find that
\begin{equation} \label{eq:sigT}
\sigma^{-2}_t (t) = \omega^2 \sum_{i=1}^N (A_i/\sigma_i)^{2} \cos^2(\omega t + \phi_i).
\end{equation}
Clearly, the accuracy of telling time depends on the relative noise strength,
{\it i.e.} the standard deviation $\sigma_i$ divided by the amplitude
$A_i$, of the respective genes, the frequency $\omega$ of the
oscillations, and the phase difference between the different
oscillations. It also depends on time, which means that the precision
with which the time can be determined, depends on the moment of
the day. The average error in the estimate, $\sigma_t(t)$ averaged over the oscillation period $T$, is
\begin{eqnarray}
\langle \sigma_t \rangle &=& \int_0^T P(t) \sigma_t(t) dt\label{meanstd}
\\
\label{sigmedio}
 &=& \frac{1}{T}\int dt \left( \sqrt{\omega^2 \sum_i^N (A_i/ \sigma_i)^{2} \cos^2(\omega t + \phi_i)}\right)^{-1}
\end{eqnarray}
It is not possible to solve this analytically, and below we have
optimized $\langle \sigma_t \rangle$ numerically. It is also of
interest to know how much the error is constant as a function of
time. To this end, we compute
\begin{equation}\label{eq:variance_error}
(\delta \sigma_t)^2 = \int_0^T P(t) (\sigma_t(t) - \langle \sigma_t \rangle)^2  dt
\end{equation}

With cross-correlations in the expressions of the downstream genes,
the off-diagonal terms of ${\bf C}$ will be non zero, which leads to
additional terms in the expression for $\sigma^2_t$. Rather than
giving the generic expression, we show the more informative expression
for $N=2$, with $x_1(t) = x(t) = A_x \sin (\omega t)$ and $x_2(t) =
y(t) = A_y
\sin(\omega t+\phi)$. The covariance matrix, which is symmetric and
semi-definite positive, is defined as
\begin{eqnarray}
\label{eq:C}
 {\bf C}& = & \left( \begin{array}{cc}
\sigma^2_x & \mbox{cov}_{xy}  \\
\mbox{cov}_{xy} & \sigma^2_y  \\
 \end{array} \right)
\end{eqnarray}
which yields for its inverse
\begin{eqnarray}
{\bf C}^{-1}& = & \frac{1}{|{\bf C}|}\left( \begin{array}{cc}
\sigma^2_y & -\mbox{cov}_{xy}  \\
-\mbox{cov}_{xy} & \sigma^2_x  \\
 \end{array} \right),
\end{eqnarray}
where the determinant is $|{\bf
  C}|=(\sigma^2_x\sigma^2_y-\mbox{cov}_{xy}^2)$. Combining this with Eq. \ref{eq:sigt} yields:
\begin{eqnarray}
 \sigma_t^{-2}(t)&=&\frac{1}{|{\bf C}|}\left[ \sigma^2_y A_x^2 \cos^2(\omega t)\right. \nonumber\\
&& - 2\mbox{cov}_{xy} A_x A_y \cos(\omega t+\phi) \cos(\omega t) \nonumber\\
&&\left.+ \sigma^2_x A_y^2 \cos^2(\omega t + \phi)\right].
\end{eqnarray}
This expression reduces to that of  Eq. \ref{eq:sigT} when the
co-variance is zero. However, in general, the error on telling time depends on the co-variance of the
fluctuations in the expression of gene $x$ and gene $y$.

The quantity $\sigma_t(t)$ is a local quantity in that it provides the
error in estimating the time as a function of the time of the
day. This quantity can be useful when certain moments of the day have
to be determined with higher precision than others. In the next
section, we discuss another quantity, the mutual information, which
makes it possible to determine how many distinct moments in time can
be specified. 
\subsubsection{Mutual Information}
The mutual information quantifies how many different input states can
be propagated uniquely \cite{Shannon1948}.
In this context, it
is defined as
\begin{equation}\label{eq:info}
I(\{x_i\};t) =\int d{\bf x} dt P(\{x_i\},t) \log \frac{P(\{x_i\},t)}{P(\{x_i\})P(t)}.
\end{equation}
The mutual information measures the reduction in uncertainty about $t$ upon measuring
$\{x_i\}$, or vice versa. The quantity is indeed symmetric in $\{x_i\}$ and $t$:
\begin{eqnarray}
I(x,y;t) &=& H(x,y) - \langle H(x,y|t) \rangle_t \label{eq:info_H_0}\\
		 &=& H(t) - \langle H(t|x,y) \rangle_{x,y} \label{eq:info_H}
\end{eqnarray}
where $H(a)= - \int da P(a) \ln P(a)$, with $P(a)$ the probability
distribution of $a$, is the entropy of variable $a$; $H(a,b|c) = -
\int da \int db P(a,b|c) \ln P(a,b|c)$ is the information entropy of
$a,b$ given $c$, with $P(a,b|c)$ the conditional probability
distribution of $a$ and $b$ given $c$, and $\langle f(c) \rangle_c$
denotes an average of $f(c)$ over the distribution $P(c)$. In our
context, Eq. \ref{eq:info_H} is perhaps the most natural expression,
since it quantifies how accurately the cell can infer the time of the
day $t$ from the expression of $x$ and $y$.

The mutual information is a global quantity, which in contrast to
$\sigma_t (t)$, does not make it possible to quantify how accurately a
given moment in time can be specified. The latter could be useful when
the system needs to change, e.g., its metabolic program at a
well-defined moment in time. On the other hand, the mutual information
does allow us to quantify how many different moments in time can be
specified, and thus how many temporal decisions the organism could
make. As Eq. \ref{eq:info_H_0} shows, the magnitude of the mutual information
depends on both $H(x,y)$ and $\langle H(x,y|t\rangle_t$. As we will
show below, cross correlations between the expression of the downstream
genes $x$ and $y$ will modify $P(x,y)$, reducing its entropy; this
tends to reduce information transmission. Yet, cross-correlations can
also decrease $\langle H(x,y|t)\rangle_t$, meaning that, on average,
the distribution of expression levels $x$ and $y$ for a given time $t$
is more narrow---a given time $t$ then maps more uniquely onto an
expression pattern $x,y$; this tends to increase the mutual
informaiton.  The balance between these two opposing factors
determines the cross correlations that maximize information
transmission.

\section{Results}

\begin{figure*}
	\centering
  \includegraphics[scale=0.7]{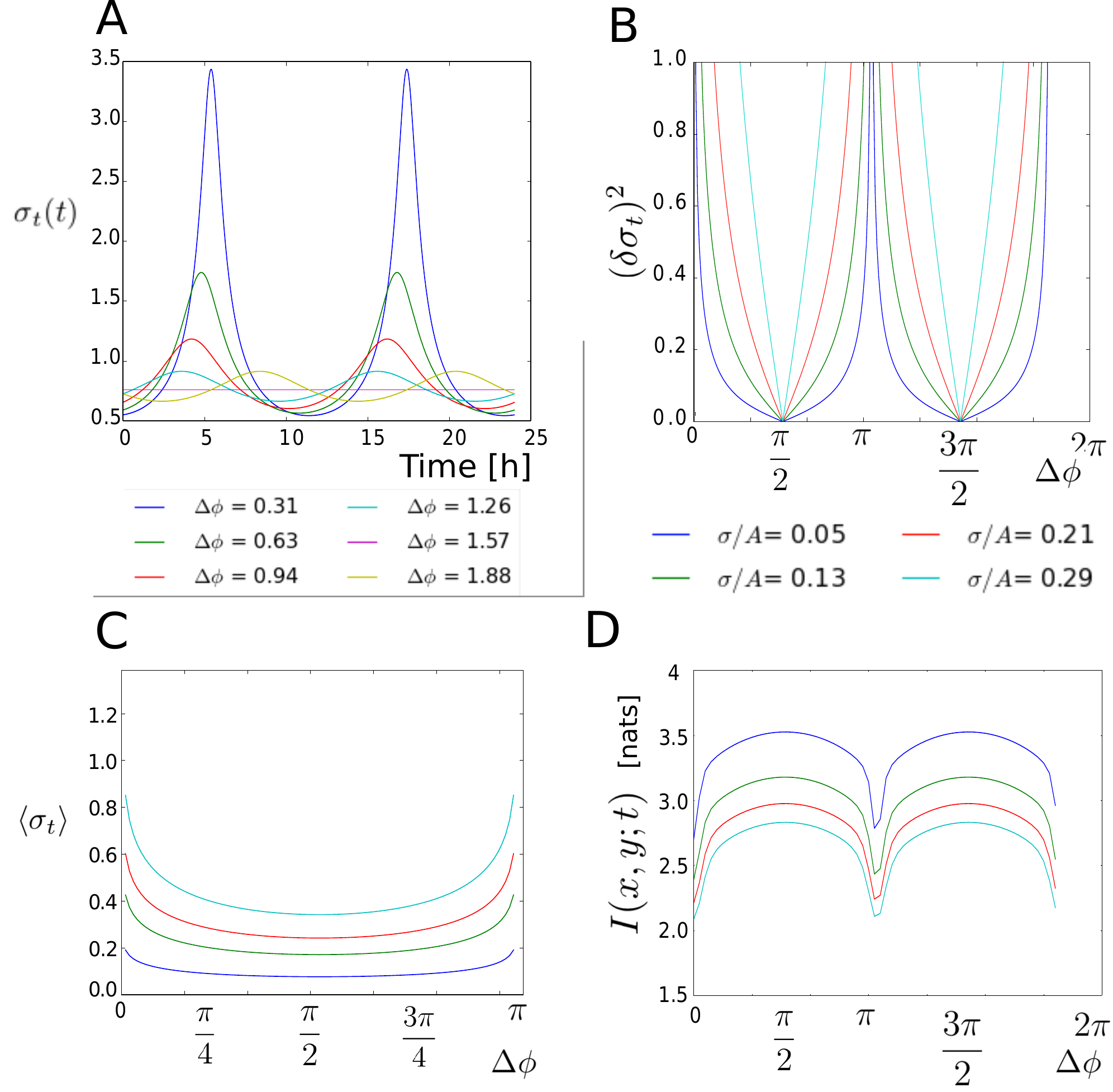}
  \caption{Estimating time via $N=2$ protein oscillations, which have
    the same relative noise strength $\widetilde{\sigma}_x =
    \sigma_x / A$. Here $A$ is the amplitude of the oscillations and
    $\sigma_x$ is the noise in the oscillations, which is here assumed
    to be constant in time, and given by the mean of the
    oscillations, $r$, taken to be the same for both oscillations;
    there are also no cross correlations. (A) The error in the estimate of time $\sigma_t (t)$
    as a function of time $t$, for different phase differences $\Delta
    \phi$ between the two oscillations. Note that for $\Delta \phi =
    \pi / 2$, the error $\sigma_t(t)$ is constant in time. (B) The
    variance $(\delta \sigma_t)^2$ in the estimate of time as a
    function of $\Delta \phi$, for different relative noise strengths
    $\widetilde{\sigma}_x$. As expected from panel A, $(\delta
    \sigma_t)^2=0$ for $\Delta \phi = \pi / 2$. (C) The mean error
    $\langle \sigma_t\rangle$ as a function of $\Delta \phi$, for
    different relative noise strengths $\widetilde{\sigma}_x$. The
    error is proportional to $\widetilde{\sigma}_x$, in accordance with
    Eq. \ref{eq:mean_error_noise_constant}. Note also that the mean
    error is minimized at $\Delta \phi=\pi/2$, although the dependence
    on $\Delta \phi$ near the optimum is weak. (D) The mutual
    information $I(x,y;t)$ between the two protein oscillations $x(t),y(t)$ and
    time $t$, for different relative noise strengths
    $\widetilde{\sigma}_x$. The mutual information increases with decreasing
    $\widetilde{\sigma}_x$, and is optimized at $\Delta \phi = \pi /
    2$. Note also that the dependence of $I(x,y;t)$ on $\Delta \phi$ is
  stronger than that of $\langle \sigma_t \rangle$ (panel C). }
  \label{fig:fig2}
\end{figure*}

\subsection{No Cross-correlations}
In this section, we consider the scenario in which there are no cross
correlations between the noise in the expression of the downstream
genes. We first study the case in which the relative noise strength,
$\widetilde{\sigma}_i\equiv \sigma_i /A_i=\widetilde{\sigma}_x$, is
the same for all genes $i$; in this scenario, we use the subscript $x$
to remind ourselves that we are considering the standard deviation in
$x$ and not in the estimate of time. We will also first assume that
$\sigma_x(t)=\sigma_x$ is constant in time, depending only on the mean
of $x$, i.e $r_x$, but not its mean instantaneous level $\bar{x}(t)$. The
latter is reasonable when the amplitude of the oscillations is small
compared to the mean.

To determine the optimal phase relation that miminizes the average
error in telling time, given by Eq. \ref{sigmedio}, we 
solve 
\begin{equation} 
\frac{d\langle \sigma_t \rangle}{d \Delta \phi_i} =0 \;\;\;\;\;\;\;\;\;\; i=1...N,
\end{equation} 
where $\Delta \phi_i = \phi_i - \phi_1$.
Setting the phase of the first oscillation to zero, i.e. $\phi_1=0$, we find that
the optimal phase relation that minimizes the average error is given
by
\begin{equation}\label{optphase}
\Delta \phi_i = (i-1) \frac{\pi}{N}  \;\;\;\;\;\;\;\;\;\; i=1...N.
\end{equation}
Clearly, in the optimal system the phases of the downstream
oscillations are evenly spaced when $\widetilde{\sigma}_x$ is the same
for all genes, and $\sigma_x$ is constant in time.

The next question is what is the phase relation that minimizes the
variance of $\sigma_t (t)$ over the oscillation period $T$,
i.e. minimizes Eq. \ref{eq:variance_error}. In the appendix we show that the
solution is also given by Eq. \ref{optphase}. Hence, the phase
relation that minimizes the average error on telling time, $\langle
\sigma_t\rangle$, is also the phase relation that minimizes the
variance of $\sigma_t (t)$. Thus, in the optimal system, the phases
are evenly spaced; this not only minimizes the average error in
telling time, but it also yields the same accuracy for all times
$t$. Moreover, for this optimal system, the average error, obtained
from Eq. \ref{eq:sigT},  is given by 
\begin{equation}
\langle \sigma_t\rangle = \frac{\widetilde{\sigma}_x T}{2\pi} \sqrt{\frac{2}{N}}
\label{eq:mean_error_noise_constant}
\end{equation}
This shows that the average error is
proportional to the relative noise strength $\widetilde{\sigma}_x=\sigma_x/A$
and inversely proportional to the square root of the number of
readout genes, $N$.

These results are illustrated in Figs. \ref{fig:fig2}A-C, for $N=2$. Panel
A shows $\sigma_t (t)$ as a function of $t$, for different phase
relations $\Delta \phi \equiv \phi_2 - \phi_1$. It is seen that, in
general, $\sigma_t (t)$, depends on $t$. However, when $\Delta \phi =
\pi / 2$, then $\sigma_t (t)$ is independent of $t$. Panel B shows
that for this phase relation, the variance $(\delta \sigma_t)^2$ is
indeed zero, while panel C shows that in this case also the average error is
minimal, in accordance with the theoretical analysis. 

Lastly, Fig. \ref{fig:fig2}D shows the mutual information $I(x,y;t)$,
obtained numerically, as a function of the phase shift, for different
noise levels. As expected, the mutual information increases as the
relative noise strength $\widetilde{\sigma}_x$
decreases. Moreover, the phase relation that minimizes the average
error, $\langle \sigma_t\rangle$, is also the phase relation that
maximizes the mutual information.

When the noise amplitude $\sigma_x$ depends on the mean instantaneous copy
number $\bar{x}(t)$ (rather than its mean averaged over the oscillation period), the noise in the output
$\sigma_x(t)$ varies in time. We will assume that $\sigma_x(t) \simeq
\sqrt{\bar{x}(t)}$, and consider as above the case that the amplitude
and the mean of the oscillations are the same for all genes,
respectively: $A_i=A_j = \dots = A$ and $r_i=r_j = \dots = r_x$. Our
analysis described in the appendix reveals that the optimal phase
relation that maximizes the mutual information and minimizes both
the variance $(\delta \sigma_t)^2$ and the mean $\langle \sigma_t\rangle$ of the error, is again given by
Eq. \ref{optphase}. However, the minimal variance, obtained for the
optimal phase relation, only reduces to zero in the limit that $r\to
\infty$; in this limit, the noise $\sigma_x(t)$ becomes constant in
time and we recover the case discussed above.  Interestingly, the average
error $\langle \sigma_t\rangle$ is larger than that in the case of
constant relative noise strength, even when the average relative noise
strength is the same.

When $N=2$ yet the relative noise strength is not the same for both
genes, $\widetilde{\sigma}_x\neq \widetilde{\sigma}_y$, the optimal
phase shift that minimizes the error and maximizes the mutual
information is again $\Delta \phi_{xy}=\pi/2$; indeed, this result,
for $N=2$, does not depend on whether $\widetilde{\sigma}$ is the same
for both genes. Also the variance $(\delta \sigma_t)^2$ is zero for
this optimal phase shift, as before.

\begin{figure*}
	\centering
  \includegraphics[scale=0.6]{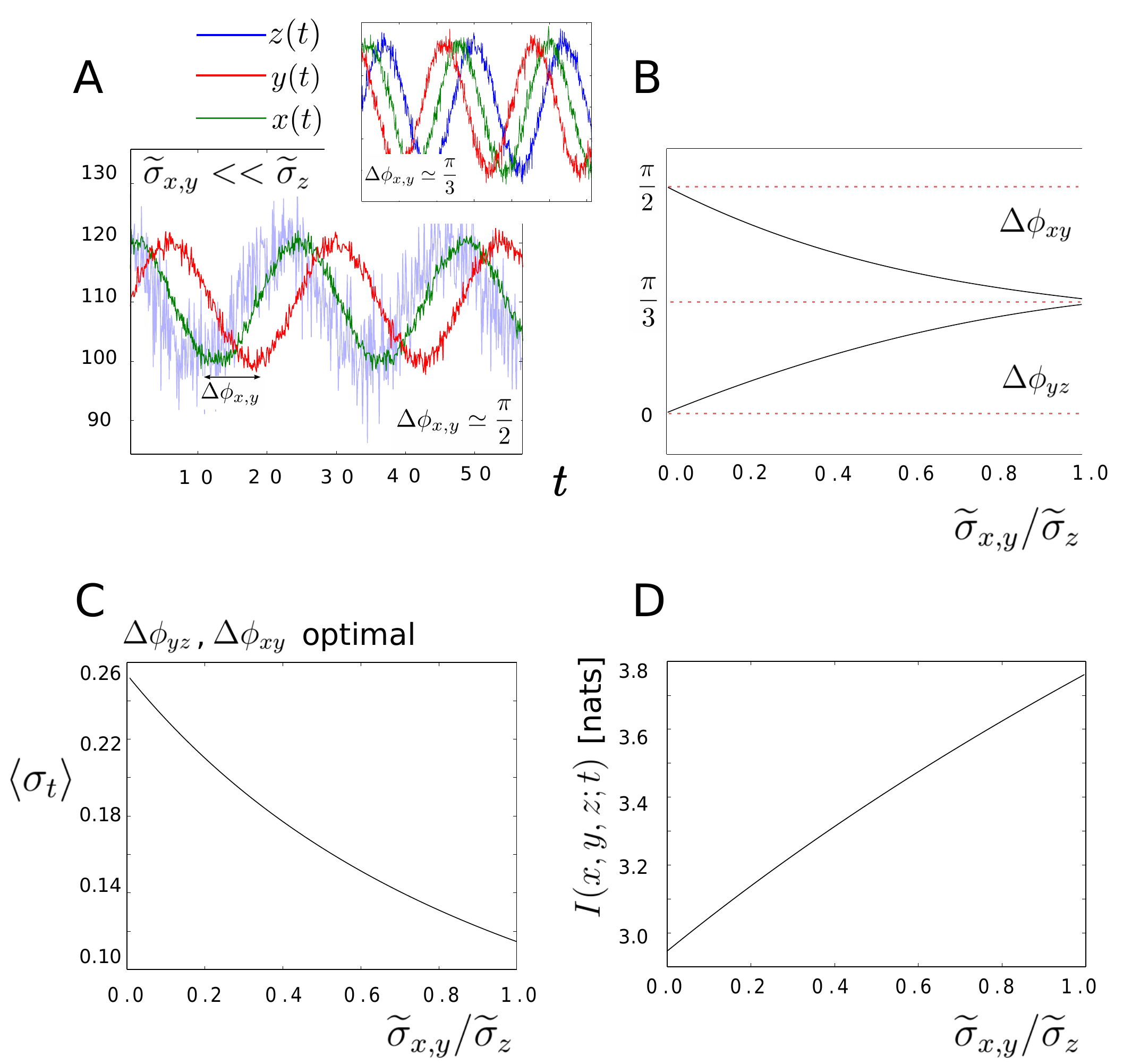}
  \caption{Estimating time via $N=3$ protein oscillations, where the
    relative noise strength $\widetilde{\sigma}_i \equiv \sigma_i / A_i$ of two oscillations is the same,
    $\widetilde{\sigma}_x = \widetilde{\sigma}_y \equiv
    \widetilde{\sigma}_{x,y}$, and different from that of the third
    oscillation, $\widetilde{\sigma}_z$. The noise $\sigma_i$ is
    assumed to be constant in time, and there are no cross
    correlations in the noise. (A) Sketch of the set up, with two
    reliable oscillations $x(t)$ and $y(t)$ and a third, more noisy
    oscillation $z(t)$. (B) The optimal phase relation that maximizes
    the mutual information $I(x,y,z;t)$ and minimizes the mean error
    $\langle \sigma_t \rangle$, as a function
    of $\widetilde{\sigma}_{x,y}/\widetilde{\sigma}_z$; here
    $\widetilde{\sigma}_{x,y}$ is kept constant while
    $\widetilde{\sigma}_z$ is varied. When
    $\widetilde{\sigma}_{x,y}/\widetilde{\sigma}_z=0$, the third gene
    $z(t)$ carries no information, and the optimal phase difference
    $\Delta \phi_{xy} = \phi_y - \phi_x$ between the oscillations of
    $x$ and $y$ is $\Delta \phi_{xy} = \pi/2$, the result for $N=2$
    oscillations; in this limit, the phase of $z$ is irrelevant. As
    $\widetilde{\sigma}_{x,y}/\widetilde{\sigma}_z$ increases, the
    third oscillation $z(t)$ becomes more important. The phase
    difference $\Delta \phi_{xy}$ between $x(t)$ and $y(t)$ decreases,
    while the phase difference $\Delta \phi_{yz}$ between
    $y(t)$ and $z(t)$ increases. When
    $\widetilde{\sigma}_{x,y}/\widetilde{\sigma}_z=1$, all genes are
    equally informative and $\Delta \phi_{xy} = \Delta \phi_{yz} =
    \Delta \phi_{zx} = \pi / 3$. (C) The mean error $\langle \sigma_t
    \rangle$ as a function
    $\widetilde{\sigma}_{x,y}/\widetilde{\sigma}_z$. It decreases as
    the third gene becomes more informative. (D) The mutual
    $I(x,y,z;t)$ increases with $\widetilde{\sigma}_{x,y}/\widetilde{\sigma}_z$. }
  \label{fig:fig3}
\end{figure*}

These results change markedly when the relative noise strength is not
the same for all genes and $N>2$. Then the optimal phase shift depends
in a non-trivial manner on $\{\widetilde{\sigma}_i\}$. The principle
is that the oscillations that contain more information about time
because they are less noisy, should be spaced further apart. More
specifically, the spacing between them should be closer to that which
maximizes the mutual information between them and time.  This
principle is illustrated in Fig. \ref{fig:fig3}A-B for three genes,
where $\widetilde{\sigma}_x=\widetilde{\sigma}_y \equiv \widetilde{\sigma}_{x, y}<
\widetilde{\sigma}_z$. Clearly, the oscillations of proteins X and Y
contain more information about time than the oscillation of protein
Z. As a consequence, the phase difference between $x(t)$ and $y(t)$,
$\Delta \phi_{xy} = \phi_y - \phi_x$, is more important in accurately
telling time than that between the two other pairs of
oscillations. The phase difference $\Delta \phi_{xy}$ is therefore
closer to $\pi / 2$, the phase difference that maximizes $I(x,y;t)$,
than those of the other pairs of genes. Indeed, the extent
to which $\Delta \phi_{xy}$ approaches $\pi/2$ depends on
$\widetilde{\sigma}_{x,y}/\widetilde{\sigma}_z$, as Fig. \ref{fig:fig3}B shows: when
$\widetilde{\sigma}_{x,y}=\widetilde{\sigma}_z$, all oscillations are equally informative and
hence the oscillations are evenly spaced, yielding $\Delta
\phi_{xy}=\Delta \phi_{yz}=\Delta \phi_{zx}=\pi/3$. In contrast, when
$\widetilde{\sigma}_{x,y}/\widetilde{\sigma}_z = 0$, $\Delta \phi_{xy}=\pi/2$, the same result
that would have been obtained if these two genes were the only ones
present. In this limit, $\widetilde{\sigma}_z$ is infinite, and $z$
carries no information on time, making its phase irrelevant.

Fig. \ref{fig:fig3}C gives the mean error $\langle \sigma_t\rangle$
and Fig. \ref{fig:fig3}D the mutual information $I(x,y,z;t)$ for the
optimal phase relation shown in panel B, as a function of
$\widetilde{\sigma}_{xy}/\widetilde{\sigma}_z$. Here, in varying
$\widetilde{\sigma}_{xy}/\widetilde{\sigma}_z$,
$\widetilde{\sigma}_{xy}$ is kept constant while
$\widetilde{\sigma}_z$ is varied between $\widetilde{\sigma}_{xy}$ and
infinity. These panels thus show the gain in employing an additional
readout protein in accurately telling time, as a function of its noise
level. The results interpolate between those for $N=2$
equally informative genes when
$\widetilde{\sigma}_{xy}/\widetilde{\sigma}_z=0$, and those for $N=3$
equally informative genes when
$\widetilde{\sigma}_{xy}/\widetilde{\sigma}_z=1$.

 \begin{figure*}[t]
 \centering
    \includegraphics[scale=0.6]{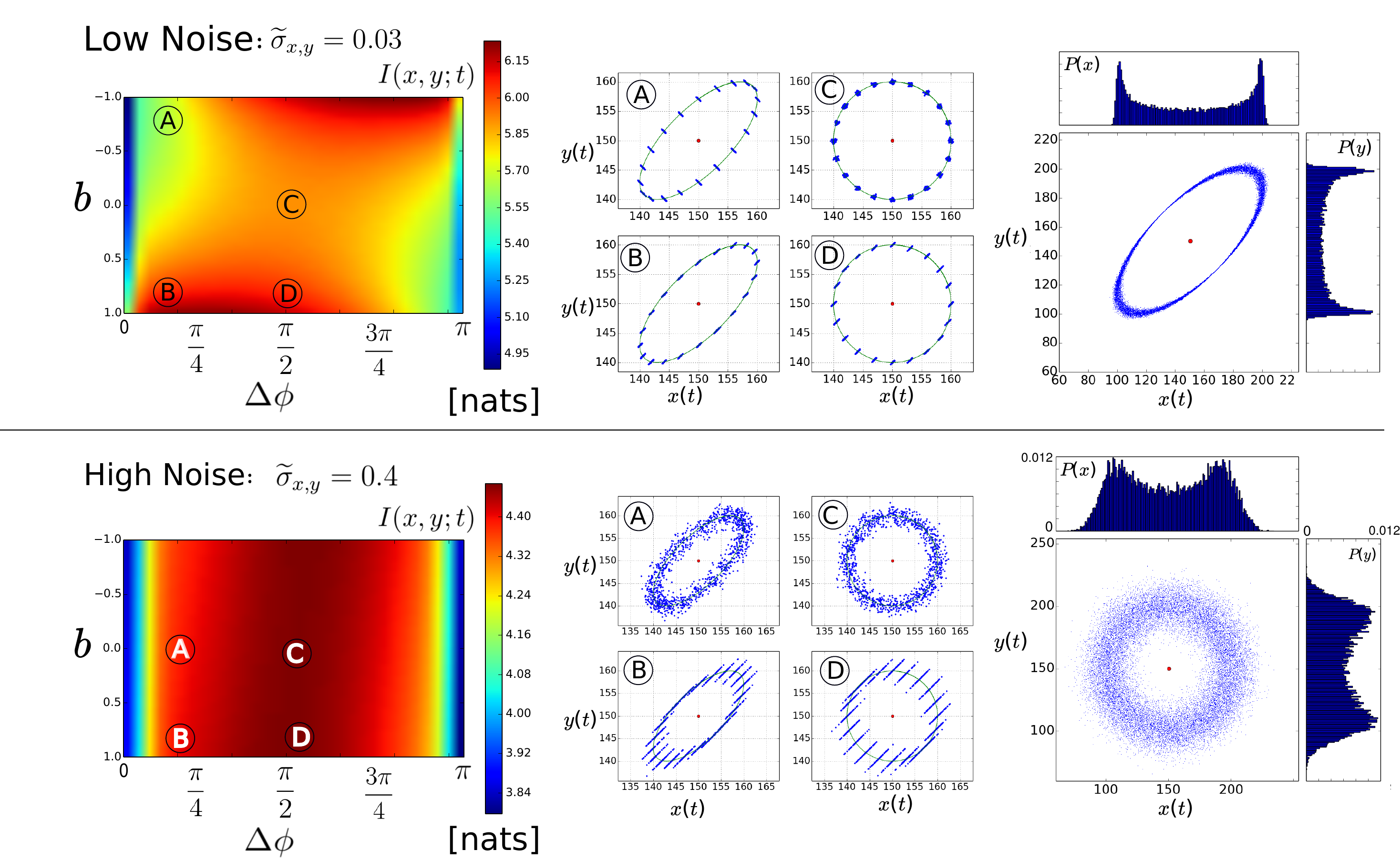}
    \caption{The importance of cross correlations between the
      fluctuations in the oscillations of the readout proteins,
      illustrated here for $N=2$ readout proteins. The top row shows
      results for the scenario in which the relative noise strength
      $\widetilde{\sigma}_i \equiv \sigma_i / A_i$ is low, while the
      bottom panel displays the results for when it is large.  In all
      cases, the relative noise strength of the two oscillations is
      taken to be the same, $\widetilde{\sigma}_x =
      \widetilde{\sigma}_y = \widetilde{\sigma}_{x,y}$. The panels in
      the left column show a heat map of the mutual information
      $I(x,y;t)$ as a function of the phase difference $\Delta \phi =
      \phi_y - \phi_x$ between the two oscillations, and the
      correlation coefficient $b$. Due to the symmetry of the problem
      the mutual information is symmetric: $I(x,y;t)_{\Delta \phi,b} =
      I(x,y;t)_{\pi - \Delta \phi,-b}$. The top-left panel shows that
      when the relative noise strength is low, the mutual information
      is maximized for $|b|\to 1$ and $\Delta \phi \neq \pi /
      2$. Cross correlations thus change the optimal phase difference,
      and more, importantly, they can enhance the mutual
      information. However, when the relative noise is large, the
      cross correlations become less important and the optimal phase
      difference approaches $\Delta \phi = \pi / 2$ (bottom left
      panel). The middle panels elucidate how cross correlations can
      affect the mutual information. Shown are, for different points
      in the heat map on the left, the average trajectory that
      ${x}(t)$ and $y(t)$ trace out during a 24 hr period (green
      solid line), with superimposed, for different times of the day,
      scatter points of $x(t)$ and $y(t)$, originating from gene
      expression noise. The main axis of the contour
      $\bar{x}(t),\bar{y}(t)$ is determined by the phase difference
      $\Delta \phi$, while the main axis of the noise (scatter points)
      is determined by the correlation coefficient $b$. There are
      moments of the day where cross correlations cause the
      distributions $P(x,y|t)$ of neighboring times $t$ to overlap
      less, thus increasing mutual information, but also moments where
      they increase the overlap, decreasing the mutual
      information. The net benefit depends on how these contributions
      are weighted. The system spends more time near the extrema of
      $\bar{x}(t),\bar{y}(t)$, as illustrated in the right panels. Consequently,
      when $\Delta \phi < \pi / 2$, positive correlations $b>0$ enhance the
      mutual information, especially when the relative noise strength
      $\widetilde{\sigma}_x$ is low (point B top row). At higher
      noise (bottom row), cross correlations are less effective in
      reducing the overlap in $P(x,y|t)$ and the phase difference
      $\Delta \phi$ becomes the dominant control parameter. }
    \label{fig:fig4}
\end{figure*}

\subsection{The importance of cross-correlations}
So far we have assumed that the noise in the expression of the
downstream genes is uncorrelated. However, in general, we expect their
noise to be correlated.  Direct or indirect regulatory interactions
between the genes can lead to correlations or anti-correlations in the
fluctuations of the protein concentrations \cite{Walczak:2010cv}. And
also noise in the input signal can lead to correlated gene
expression. In fact, the extrinsic contribution to the noise in gene
expression is often larger than the intrinsic one
\cite{Taniguchi:2010cb}, which can induce pronounced correlations
between the expression of the downstream genes. Intuitively, we may
think that if we need to infer an input variable $t$ from two output
variables $x$ and $y$, then cross-correlations between $x$ and $y$
reduce the accuracy of the estimate---asking two persons $x$ and $y$
a question about $t$ seems to give more information when $x$ and $y$
give independent answers. However, this intuition is not always
correct, as will become clear. Indeed, in this section
we study how correlations between the expression of downstream genes
affect the precision by which cells can tell time.

 \begin{figure*}[t]
 \centering
    \includegraphics[scale=0.9]{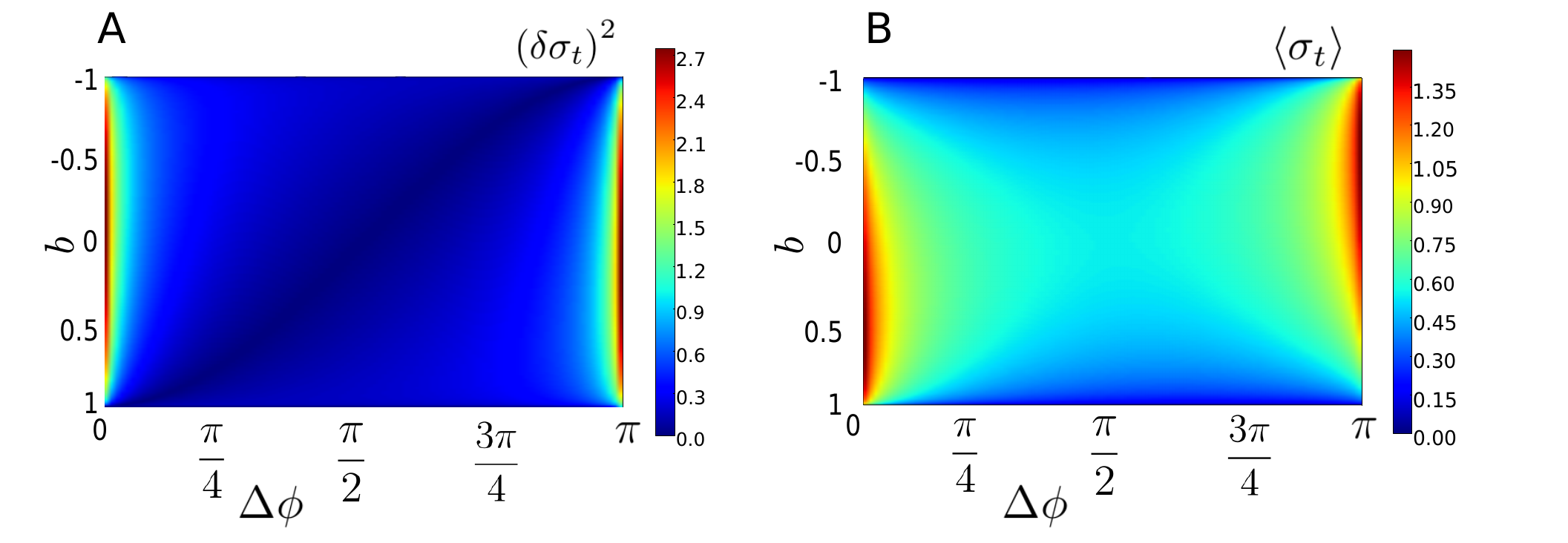}
    \caption{Importance of cross correlations in reducing the error in
      estimating the time, as estimated from $N=2$ protein
      oscillations. Heat maps of the the variance in the error
      $(\delta \sigma_t)^2$ (A) and the mean error $\langle \sigma_t
      \rangle$ 
      (B), as a function of the phase difference $\Delta \phi$ between
      the two oscillations and the correlation coefficient $b$ of the
      fluctuations in the oscillations. The relative noise strength
      $\widetilde{\sigma}_x$ is the same for both oscillations, and
      equal to that of the low-noise scenario in Fig. \ref{fig:fig4},
      $\widetilde{\sigma}_x = 0.03$. It is seen that cross
      correlations can reduce the mean error. Comparing against the
      top-left panel of Fig. \ref{fig:fig4} shows, however, that the
      positions of the optima are different for the two quantities,
      the mean error $\langle \sigma_t \rangle$ and the mutual
      information $I(x,y;t)$ , respectively. This is because the
      quantities $\sigma_t(t)$ (Eq. \ref{sigmedio}) and $H(t|x,y)$
      (Eq. \ref{eq:info_H_2}) are averaged over different
      distributions, the uniform distribution $P(t)$ and the
      non-uniform distribution $P(x,y)$, respectively. }
    \label{fig:fig5}
\end{figure*}

In order to dissect the effect of cross-correlations, we study two
downstream genes, $N=2$, and take both the amplitudes of their
oscillations and their expression noise to be equal: $A_x = A_y = A$,
$\sigma_x=\sigma_y=\sigma_{x,y}$, respectively. Using the latter, we can renormalize the
covariance matrix Eq. \ref{eq:C}:
\begin{equation}
{\bf C}=  \left( \begin{array}{cc}
\sigma_x & \mbox{cov}_{xy}  \\
\mbox{cov}_{xy} & \sigma_y  \\
 \end{array} \right) = \sigma_{x,y} \left( \begin{array}{cc}
1 & b  \\
b & 1  \\
 \end{array} \right),
\end{equation} 
where $b$ is the correlation coefficient, denoting the
cross-correlation strength: $b=1$ implies that the noise in the
expression of X and Y is fully correlated, while $b=-1$ implies full
anti-correlation. We computed numerically how $I(x,y;t)$, $\langle
\sigma_t\rangle_t$ and $(\delta t)^2$ depend on the phase shift
$\Delta \phi = \phi_y - \phi_x$, the relative noise strength
$\widetilde{\sigma}_{x,y} = \sigma_{x,y} / A$, and the correlation
coeffient $b$.

Fig. \ref{fig:fig4} shows the mutual information $I(x,y;t)$ as a
function of $\Delta \phi$ and $b$, both for low noise, with
$\widetilde{\sigma}_{x,y} = 0.03$ (panels top row), and high noise, with
$\widetilde{\sigma}_{x,y} = 0.4$ (panels bottom row).  The following points
are worthy of note. First, as expected, $I(x,y;t)$ is symmetric with
respect to $\Delta \phi$ and $b$: $I(x,y;t)_{\Delta \phi,
  b}=I(x,y;t)_{2\pi - \Delta \phi,-b}$. Secondly, depending on the
phase shift $\Delta \phi$, correlations ($b>0$) or
anti-correlations ($b<0$) can enhance the mutual information,
especially when  the relative
noise strength $\widetilde{\sigma}_{x,y}$ is low (top panel).
Concomittantly, the optimal phase shift $\Delta \phi$ that maximizes the mutual
information depends on the cross correlation $b$. At low noise, the
mutual information is maximized either at $0<\Delta \phi^*<\pi / 2$
and $b\approx 1$ or at $\pi - \Delta \phi^*$ and $b\approx -1$. At high
noise, cross correlations no longers help to improve the mutual
information  (bottom panel). Moreover, the optimal phase shift is at $\Delta \phi^* \approx
\pi / 2$. We now discuss the origin of these observations.

To elucidate these observations, we start from the definition of the
mutual information (see Eq. \ref{eq:info_H}):
\begin{equation}
I(x,y;t) = H(t) - \langle H(t|x,y) \rangle_{x,y}\label{eq:info_H_2}
\end{equation}
Here, $H(t)$ is the entropy of the input signal, with $P(t) = 1
/T$. It does not depend on the design of the downstream readout
system. In contrast, the second term, $\langle H(t|x,y)
\rangle_{x,y}$, does depend on it. We now describe how changing
$\Delta \phi$ and $b$ affects this term, using the scatter plots and
distributions in the middle and right column of Fig. \ref{fig:fig4}.

The middle panel shows for different combinations of $b$ and $\Delta
\phi$, corresponding to the points A,B,C,D in the heat map of
$I(x,y;t)$ (left panel), scatter plots of $x(t)$ and $y(t)$.  The
overall shape of each scatter plot is determined by the phase
difference $\Delta \phi$. When $\Delta \phi = \pi / 2$ (points C and
D), the average expression levels $\bar{x}(t)$ and $\bar{y}(t)$ trace
out a circle in state space during a 24 hr period, while when $\Delta
\phi = \pi / 4$ (points A and B), they carve out an ellipsoidal path;
these mean paths are indicated by thin solid green lines in the
scatter plots. For each moment of the day, however, $x$ and $y$ will
exhibit a distribution of expression levels, due to gene expression
noise. This distribution $P(x,y|t)$ is shown as scatter points $(x,y)$
for different yet evenly spaced times $t$ in the respective subpanels.  When
the main axis of $P(x,y|t)$ is perpendicular to the local tangent of
the mean path of $\bar{x}(t),\bar{y}(t)$, then cross correlations
reduce $H(t|x,y)$ for that period of the day: the cross correlations
cause the distributions $P(x,y|t)$ for neighboring times $t$ to
overlap less, meaning that a given point $(x,y)$ maps more uniquely
onto a given time $t$. This tends to increase the mutual
information. However, as the middle panel illustrates, there are not
only moments of the day when the main axis of the scatter points is
perpendicular to the local tangent of the mean path, but also times
when they are parallel, in which case cross correlations are
detrimental. Whether the net result of cross correlations is
beneficial, depends on how these different contributions are weighted:
$H(t|x,y)$ has to be averaged over $P(x,y)$, see
Eq. \ref{eq:info_H_2}. When $\Delta \phi = \pi / 2$, the mean path
$\bar{x}(t),\bar{y}(t)$ is circular, yet the net effect of
correlations on the mutual information is already positive (left
panel), and independent of the sign of $b$. For $\Delta \phi \neq \pi
/ 2$, the effect depends on the sign of $b$. Moreover, the effect is
also stronger, because the
system spends more
time near the extrema of $\bar{x}(t),\bar{y}(t)$, as the right panel
illustrates.  When $\Delta \phi < \pi / 2$, positive correlations in
the expression of $x$ and $y$ ($b>0$) cause the main axis of
$P(x,y|t)$ to be perpendicular to the local tangent of
$\bar{x}(t),\bar{y}(t)$ near the extrema (point B), thus increasing
the mutual information, while anti-correlations ($b<0$) cause
$P(x,y|t)$ to be parallel to it (point A), decreasing the mutual
information. For $\Delta \phi \rightarrow \Delta \phi - \pi/2$
precisely the oppositive behavior is observed, because the mean path
of $\bar{x}(t), \bar{y}(t)$ (the ellipse) is flipped vertically.  The
principal observation is thus that cross-correlations can enhance the
mutual information by allowing for a less overlapping tiling of state
space, and hence a less redundant mapping between the input $t$ and
output $(x,y)$.
 
For higher noise (panels in lower row of Fig. \ref{fig:fig4}), each $P(x,y|t)$
becomes wider, which means that the benefit of introducing cross
correlations in reducing the overlap between different $P(x,y|t)$
(corresponding to different times $t$), decreases. Indeed, at higher
noise, the mutual information depends much more weakly on the
magnitude of the cross correlations (left panel bottom row). The key
control parameter is now the phase shift $\Delta \phi$. For $\Delta
\phi = \pi / 2$, the distributions $P(x,y|t)$ are most evenly spaced. This
minimizes the overlap between them and maximizes the mutual
information.

Fig. \ref{fig:fig5} shows the the variance in the error, $(\delta
t)^2$, and the average error in telling time, $\langle
\sigma_t \rangle$, as a
function of $\Delta \phi$ and $b$, for $\widetilde{\sigma}=0.03$ (as
in the top row of Fig. \ref{fig:fig4}). It is seen that increasing
correlations $|b|$ can reduce the average error. Surprisingly,
however, for $|b|\approx 1$, the average error $\langle \sigma_t
\rangle$ is minimized at a phase shift that does not maximize the
mutual information, as a comparison with Fig. \ref{fig:fig4} shows. 
This is because of how the respective quantities are
averaged. The quantity $\sigma_t (t)$ is averaged over $P(t)$, which
is uniform in time, while $H(t|x,y)$ is averaged over $P(x,y)$, which gives
more weight to those points $(x,y)$ that are more probable.

\section{Discussion}
Cells can increase the transmission of temporal information by
increasing the number of oscillatory signals $N$ used to infer the
time. In the analysis presented here, it is assumed that the system is
linear and obeys Gaussian statistics, yet, especially at high noise,
it might be beneficial to use non-linear input-output relations to
enhance information transmission \cite{DeRonde:2014fq}. Nonetheless,
our linear model highlights that this is a rich problem. The precision
of telling time depends on the relative noise $\widetilde{\sigma}_i =
\sigma_i / A_i$ of the oscillatory signals, their phase shift, and the
cross-correlations between them. When the relative noise
$\widetilde{\sigma}_i$ is the same for all genes, the optimal phase
relation that maximizes the mutual information and minimizes the error
is one in which the phases are spaced evenly. Under this condition,
the error in telling time is also uniform in time, provided that the
noise $\sigma_i(t)$ is constant in time, which, to a good
approximation, is the case when the amplitude of the oscillations is
large compared to the mean. This is akin to what has been observed for
the fruitfly {\it Drosophila}, where the expression pattern of the gap
genes allows the nuclei to specificy their position with nearly
uniform precision along the anterior-posterior axis \cite{Dubuis2013}.
When the relative noise amplitudes $\widetilde{\sigma}_i$ are not the
same for all signals, then the design principle for maximizing
information transmission is that the oscillatory signals which are
more reliable, should be spaced more evenly.  Lastly, we have
addressed the role of cross correlations between the fluctuations in
the oscillatory signals. When the relative noise is large,
cross-correlations do not significantly affect information
transmission. However, the situation changes markedly in the low-noise
regime. In this regime, cross-correlations change the optimal phase
shift that maximizes information transmission. More strikingly, they
can increase the mutual information. At low noise, cross correlations
can thus reduce the error in telling time and enhance the transmission
of temporal information. This phenomenon is similar to what has been
observed for neural networks \cite{Tkacik2010} and spatial gene
expression patterns during embryonic development, where
cross-regulatory interactions between genes can enhance the precision
by which cells or nuclei determine their spatial position within
the developing embryo \cite{Tkacik:2009ta,Walczak:2010cv,Dubuis2013}. In
all these cases the principle is that cross-correlations make it
possible to tile the output space more efficiently, thus allowing for
a less redundant input-output mapping. This is particularly important
when the noise is low, and noise averaging is not important, but
efficient tiling of state space is
\cite{Tkacik:2009ta,Walczak:2010cv}.

The question that remains is how cells can optimize the relative noise
of the oscillatory signals, their phase difference and their
cross-correlations. Fluctuations in the input will lead to correlated
fluctuations in the oscillations of the output components. Our
analysis shows that these correlations can be beneficial. Moreover,
they can be tailored via cross-regulatory interactions between the
target genes downstream, as in the gap-gene system of {\it Drosophila}
\cite{Tkacik:2009ta,Walczak:2010cv,Dubuis2013}. Here, it should be
realized that in our analysis we assume that the noise is uncorrelated
from the signal; indeed, the mean trajectory $(\bar{x}(t),\bar{y}(t))$
does not depend on the noise. Cross-regulatory interactions will,
however, not only affect the noise and hence $P(x,y|t)$, but also the
mean trajectory $(\bar{x}(t),\bar{y}(t))$. This will not change the principle
that noise correlations can enhance the input-ouput mapping, but it
will influence the magnitude of the effect. On the other hand,
extrinsic noise sources such as the availability of ribosomes, may
lead to correlated fluctuations in the expression of $x(t)$ and
$y(t)$, while leaving their mean unchanged, as assumed
here. Experiments will have to tell whether cells use noise
correlations to enhance the precision of telling time. The
cyanobacterium {\it S. elongatus} is arguably the best model system to
test these ideas. It will certainly be of interest to investigate
whether {\it S. elongatus} exploits cross-regulatory interactions
between the genes downstream from RpaA to enhance its information on
time.

The relative noise of the oscillations depends on the noise $\sigma_i$
and the amplitude $A_i$ of the oscillations.  The contribution from
the intrinsic noise is expected to scale with the copy number $X$ as
$\sigma_{{\rm in},i} \sim \sqrt{X_i}$, which, if the amplitude is
small compared to the mean $r_i$, means that $\sigma_{{\rm in},i} \sim
\sqrt{r_i}$. The relative intrinsic noise thus goes as
$\widetilde{\sigma}_{{\rm in},i} \sim \sqrt{r_i} / A_i$. For the model
presented in section \ref{sec:Model}, it is given by
\begin{eqnarray}
\widetilde{\sigma}_{{\rm in},i} &\simeq& \sqrt{r_i} / A_i\\
&=&  \sqrt{r_s/
  f_i}\sqrt{(\mu_i^2 + \omega^2)/\mu_i} / A_s.
\end{eqnarray} 
Clearly, the relative noise strength $\widetilde{\sigma}_{{\rm in},i}$
decreases with $A_s$: the amplitude of the oscillations of the readout
is proportional to that of the input. The relative noise strength
decreases with the square root of $f_i$, because the gain $f_i$
increases not only the amplitude of the output oscillations, $A_i
\propto f_i$, but also their mean $r_i$ and thereby the noise,
$\sigma_{{\rm in},i} \propto \sqrt{r_i} \propto \sqrt{f_i}$. It
increases with the mean $r_s$ of the input oscillations, because that
increases the mean $r_i$ of the output oscillations and thereby the
noise $\sigma_{{\rm in},i}$, but not their amplitude, thus decreasing
the relative noise strength $\sigma_{{\rm in},i} / A_i$. Finally,
there exists an optimal protein decay rate $\mu_{\rm opt}=\omega$ that
minimizes the relative noise strength and hence maximizes information
transmission. This optimum arises from a trade-off between the
amplitude of the signal and the intrinsic noise: for $\mu \gg \omega$,
increasing $\mu$ reduces the gain and hence the amplitude $A_i$ as
$A_i \propto 1/\mu$ (Eq. \ref{eq:Amp}) while the noise decreases
more slowly as $\sqrt{r_i}\propto 1/\sqrt{\mu}$, thus increasing the relative noise
strength $\widetilde{\sigma}_{{\rm in},i}$; in contrast, for $\mu \ll
\omega$, the amplitude $A_i$ becomes independent of $\mu_i$
(Eq. \ref{eq:Amp}) while the
noise continues to rise as $\mu_i$ decreases, thus again increasing
the relative noise strength.

For the transmission of a fluctuating input signal, a similar
trade-off between the gain and the intrinsic noise has been observed
in \cite{tostevin10} and a related trade-off between mechanistic error
arising from the intrinsic noise and dynamical error due to the
distortion of the input signal has been described in
\cite{Bowsher:2013jh}. A seemingly similar but distinct trade-off,
also leading to an optimal decay rate of the output component, has
been reported in \cite{Becker2015}: in that study the optimal decay
rate arises from the trade-off between tracking the input signal and
integrating out the noise in the input signal. Indeed, in our
discussion here, we have so far ignored the extrinsic noise in the
input signal, and only focused on the intrinsic noise. However, the
decay rate $\mu_i$ does
not only affect the output copy number and thereby the intrinsic
noise, it also determines how effectively fluctuations in the input
signal can be integrated out. More
specifically, if the noise in the input $\xi_s$ (Eq. \ref{eq:eta}) is
independent from the input signal, has amplitude $\sigma_s$ and decays
exponentially with correlation time $\lambda$, then we expect that the
extrinsic contribution to the output noise is $\sigma^2_{{\rm ex},i} =
g_i^2 \mu_i / (\mu_i+\lambda) \sigma^2_s$
\cite{paulsson2003,tanase-nicola06}, where the gain is
$g_i=f_i/\mu_i$. Hence, the relative extrinsic noise is
\begin{equation}
\sigma_{{\rm
    ex},i}/A_i=1/\mu_i\sqrt{(\mu_i^2+\omega^2)\mu_i/(\mu_i+\lambda)}\sigma_s.
\end{equation}
 We
first note that, in contrast to the relative contribution of the
intrinsic noise, $\sigma_{{\rm in,i}}/A_i$, the relative extrinsic noise
does not depend on $f_i$: increasing $f_i$ raises not only the
amplitude of the signal, but also that of the noise; increasing $f_i$ is
thus only useful in raising the signal above the {\em intrinsic} noise. Secondly,
for $\mu_i \gg \omega,\lambda$, $\sigma_{{\rm ex},i}/A_i \simeq
\sigma_s$, because the time integration factor $\mu_i /
(\mu_i+\lambda)$ becomes constant (independent of $\mu_i$), and both
the amplitude of the signal, $A_i$, and the amplification of the input
noise, $g_i$ decrease as $\mu_i^{-1}$. For $\mu_i \ll \omega,
\lambda$, $\sigma_{{\rm ex},i}/A_i \simeq \omega \sigma_s
/\sqrt{\mu_i\lambda}$, because the amplitude $A_i$ becomes independent
of $\mu_i$, while the extrinsic contribution $\sigma_{{\rm ex},i}$
rises with decreasing $\mu_i$ as $1/\sqrt{\mu_i}$. In fact, the
relative strength of the extrinsic noise $\sigma^2_{{\rm ex},}/A_i$
has a minimum at $\mu_{\rm ex}^{\rm opt} = (\omega^2 / \lambda)
(1+\sqrt{1+(\lambda/\omega)^2})$. We thus conclude that both the
relative strength of the intrinsic and extrinsic noise exhibit a
minimum as function of $\mu_i$, meaning that there is an optimal
protein lifetime that maximizes information transmission. 

Lastly, how could cells optimize the phase relation between the oscillations of
the readout proteins? In the simple model of \ref{sec:Model} there is
only one control variable, namely the protein degradation rate
(Eq. \ref{eq:phase}). Clearly, it is not possible, in general, to
simultaneously set the decay rate such that the relative noise
strength is minimized, as described above, and the phase difference is
optimized. However, the simple model of \ref{sec:Model} ignores that
gene expression is, in fact, a multi-step process leading to a delay,
and it is possible that nature has tuned this delay so as to optimize
the phase relation between the output oscillations. In addition, cells
could use gene expression cascades to adjust the delay. Whether cells
employ these mechanisms to optimize the phase relation is an interesting
question for future work.

\appendix
\section{The optimal phase relation in the absence of cross
  correlations}
We would like to compute the phase relation that minimizes the
variance of the error, $(\delta \sigma_t)^2$, as given by
Eq. \ref{eq:variance_error}, in the absence of cross correlations. However, the problem is that
Eq. \ref{eq:sigt} is an expression for $\sigma_t^{-2}(t)$, not
$\sigma_t(t)$. Hence, while it is fairly straightforward to
derive the variance of $\sigma_t^{-2}$, i.e. $\langle
(\sigma_t^{-2})^2\rangle - \langle \sigma_t^{-2}\rangle^2$, it is
impossible, in general, to derive analytically the variance of the
quantity we are interested in, $(\delta \sigma_t)^2=\langle
\sigma_t^2\rangle - \langle \sigma_t\rangle^2$. However, we know that
if the variance of a function $g(t)$ is zero, $\sigma^2_g = \int_0^T
dt P(t) (g(t) - \langle g(t) \rangle)^2=0$, and $g(t)$ is thus a
constant (independent of time), that then a) $\langle f(g(t))\rangle =
f(\langle g(t) \rangle)$ and b) the variance of $f(t)=f(g(t))$ is zero,
$\sigma^2_f = \langle f^2 \rangle - \langle f\rangle^2 = 0$. We now apply this logic with the
identification $g(t) = \sigma_t (t)$ and $f(t)=g^{-2}(t)$. The trick
that we thus employ is to establish that the variance which we can
compute, $\sigma^2_f = \langle (\sigma_t^{-2})^2\rangle - \langle
\sigma_t^{-2}\rangle^2$, is zero. If this is true, then we know that a)
the variance of the quantity that we are interested in, $\sigma^2_g = (\delta
\sigma_t)^2$, must be zero as well. Moreover, we then also know that b)
$\langle \sigma_t \rangle = \sigma_t = 1/\sqrt{\langle \sigma^{-2}_t (t)\rangle}$.

There are two points worthy of note. First, as mentioned, above, when
$\sigma^2_f = \langle (\sigma^{-2}_t (t)) ^2\rangle - \langle
\sigma^{-2}_t (t) \rangle^2 = 0$, then $(\delta \sigma_t)^2=0$. In
this case, the phase relation that minimizes $\sigma^2_f$ is the phase
relation that minizes $\delta \sigma^2_t$ (making it
zero indeed). However, when $\sigma^2_f \neq 0$, then the phase
relation that minimizes $\sigma^2_f$ is {\em not} necessarily the phase
relation that minimzes $\delta \sigma^2_t$. Secondly, the phase
relation that minimizes $(\delta
\sigma_t)^2$, is not necessarily the phase relation that minimizes
$\sigma_t$, {\em even when $(\delta \sigma_t)^2=0$}. We need to check
either numerically or, if possible, by analytically minizing $\langle
\sigma_t \rangle$ whether this is true or not. The same holds for the
mutual information: the phase relation that minimizes  $(\delta
\sigma_t)^2$, is not necessarily the phase relation that maximizes the
mutual information.

\subsection{The phase relation that minimizes $(\delta \sigma_t)^2$
  when the relative noise strengths are the same}

As explained above, to obtain the optimal phase relation that makes
$(\delta \sigma_t)^2=0$, we aim to find the phase distribution for which:
\begin{equation}
\sigma^2_f = \langle (\sigma^{-2}_t (t)) ^2\rangle - \langle \sigma^{-2}_t (t) \rangle^2 = 0.\label{eq:var_f}
\end{equation}
When the cross correlations are zero, $\sigma^{-2}_t (t)$ is given by
Eq. \ref{eq:sigT}. The second term in the expression above, $\langle
\sigma^{-2}_t (t) \rangle^2$, is then, for the case that the noise and
the ampltidues are the same for all genes, given by
\begin{equation}
\langle \sigma^{-2}_t (t) \rangle = \left(\frac{ 2\pi A}{\sigma_x T}\right)^2 \frac{N}{2}.\label{eq:mean_sigmainvsq}
\end{equation}
The first term in Eq. \ref{eq:var_f} can be obtained recursively, and
is given by 
\begin{eqnarray}
&&\langle (\sigma^{-2}_t (t)) ^2\rangle = \nonumber\\ 
 && K \left[ \frac{N(2N+1)}{8}+ \frac{1}{4}\sum_{i<j}^N \cos(2(\phi_i
   -\phi_j))\right]\label{eq:var_f_ft_const_noise}
\end{eqnarray}
where $K$ is a constant, $K =\left(\frac{ 2\pi A}{\sigma_x T}\right)^4$.
As expected this quantity depends on the phase relation.

Instead of finding the phase relation that makes the difference
between the two terms of
$\sigma^2_f$ in Eq. \ref{eq:var_f} zero, we now want to find the
relation that makes the ratio of the two terms unity, which is
equivalent, but mathematically more convenient. This yields 
\begin{equation}\label{cond1}
\frac{2}{N} \sum_{i \le j}^N \cos(2(\phi_i-\phi_j)) = -1.
\end{equation}

By solving this as a function of $N$, we can recognize a pattern,
which reveals that the optimal phase relation that minimizes $(\delta
\sigma_t)^2$ is given by 
\begin{equation}\label{phasedif}
\phi_i - \phi_j = \frac{\pi}{N}(i-j). 
\end{equation}
This means that the $i$-th signal has  a phase $\Delta \phi_i =
(i-1) \frac{\pi}{N}$, as found for the phase relation that minimizes
$\langle \sigma_t \rangle$, given by Eq \ref{optphase}. So in the case
where the correlations are zero, the optimal
phase shift minimizes both $\langle \sigma_t \rangle$ and its
variance. Moreover, the mean error $\langle \sigma_t \rangle$ can then
directly be obtained from Eq. \ref{eq:mean_sigmainvsq}.

\subsection{The phase relation that minimizes $(\delta \sigma_t)^2$
  when the noise $\sigma_x$ is not constant in time}
We now consider the case that $\sigma_i \simeq \sqrt{\bar{x}_i(t)}$,
which means that $d\sigma_i / dt \neq 0$. In order to highlight the
role of the time-varying noise, we keep $A_i = A_j = \dots = A$, $r_i
= r_j = \dots = r$. The variance of $\sigma^{-2}(t)$ is given by:
\begin{eqnarray}
\sigma^2_f &=& \langle (\sigma^{-2}_t (t)) ^2\rangle - \langle
\sigma^{-2}_t (t) \rangle^2 \nonumber\\
& = &\left(\frac{A^3 (2\pi)^2}{16 T^2}\right)^2\{ N
+2N(N+1)r^2 + \nonumber\\
& &\sum_{i \le j}^N  \left[\cos(\phi_i - \phi_j) +4r^2 \cos[2(\phi_i -\phi_j)\right] +\nonumber \\
& &\left. 4 r^2 \cos(\phi_i -\phi_j)\right]\} - \left (\frac{NA^3r (2\pi)^2}{2T^2}\right)^2\label{eq:var_f_sigma_not_constant}
\end{eqnarray}

We note that this expression, in contrast to that for the case in which
$\sigma_i$ is constant in time, depends on the mean expression level
of $x$, $r$. We find numerically that the phase relation that
minimizes $\sigma^2_f$ is the same as that for the scenario in which
$\sigma_i$ is constant in time, Eq. \ref{phasedif}. However,
$\sigma^2_f$ and hence $(\delta \sigma_t)^2$ are only zero, when $r
\to \infty$. We also find numerically that the phase relation that
minimizes $\sigma^2_f$ equals the phase relation that minimizes the mean
error $\langle \sigma_t \rangle$ and maximizes the mutual information.

\subsection{The phase relation that minimizes $(\delta \sigma_t)^2$
  when the relative noise strengths are {\em not} the same}
To assess the importance of differences in the relative noise
strength, we will assume again that $\sigma_i(t) = \sigma_i$ is
constant in time. Defining the relative noise {\em amplitude}
$\widetilde{A}_i \equiv \widetilde{\sigma}_i^{-1} \equiv A_i / \sigma_i$,
the variance of $\sigma^{-2}(t)$ is given by:
\begin{eqnarray}
&&\sigma^2_f = \langle (\sigma^{-2}_t (t)) ^2\rangle - \langle
\sigma^{-2}_t (t) \rangle^2 \nonumber\\
&&= \frac{1}{8}\left(\frac{2\pi}{T}\right)^4 \left[\sum_{i=1}^N 3\widetilde{A}_i^2 +\sum_{i \le j}^N \left(\left[4+2\cos[2(\phi_i-\phi_j)] 
\widetilde{A}_i \widetilde{A}_j\right]\right)\right] \nonumber\\
 && -\left( \frac{1}{2}\left(\frac{2\pi}{T}\right)^2 \sum_{i=1}^N \widetilde{A}_i\right)^2
\label{eq:var_f_sigma_diff_noise}
\end{eqnarray}
It can
be verified that this reduces to Eq. \ref{eq:var_f} when $\sigma_i /
A_i$ is the same for all genes. Following the
logic applied for that scenario, we find that the optimal phase
relation that makes $\sigma^2_f=0$ is given by

\begin{eqnarray}\label{ph_sigdif}
\sum_{i \le j}^N \cos\left[ 2(	\phi_i -\phi_j)  \widetilde{A}_i \widetilde{A}_j \right] \widetilde{A}_i^2 \widetilde{A}_j = \nonumber\\  
\sum_{i,j =1}^N  \widetilde{A}_i \widetilde{A}_j -\frac{1}{2}\sum_{i =1}^N 3 \widetilde{A}_i^2  - 2\sum_{i \le j}^N \widetilde{A}_i \widetilde{A}_j 
\end{eqnarray}
This expression reduces to Eq. \ref{cond1} when $\sigma_i /
A_i$ is the same for all genes. It can be verified numerically that
the phase relation that makes $\sigma^2_f$ and hence $(\delta
\sigma_t)^2$ zero, is also the phase relation that minimizes the
mean error $\langle \sigma_t\rangle$ and maximizes the mutual information.

\begin{acknowledgments}
  We thank Giulia Malaguti for a critical reading of the
  manuscript. This work is part of the research programme of the
  Foundation for Fundamental Research on Matter (FOM), which is part
  of the Netherlands Organisation for Scientific Research (NWO).
\end{acknowledgments}

 \bibliographystyle{unsrt}

     \bibliography{paper_info.bib}

\begin{thebibliography}{10}

\bibitem{Johnson2008}
Carl~Hirschie Johnson, Martin Egli, and Phoebe~L Stewart.
\newblock {Structural insights into a circadian oscillator.}
\newblock {\em Science}, 322(5902):697--701, October 2008.

\bibitem{Liu:1997uv}
C~Liu, D~R Weaver, S~H Strogatz, and S~M Reppert.
\newblock {Cellular construction of a circadian clock: period determination in
  the suprachiasmatic nuclei}.
\newblock {\em Cell}, 91(6):855--860, 1997.

\bibitem{Yamaguchi:2003jj}
S~Yamaguchi.
\newblock {Synchronization of Cellular Clocks in the Suprachiasmatic Nucleus}.
\newblock {\em Science}, 302(5649):1408--1412, November 2003.

\bibitem{Mihalcescu:2004ch}
I~Mihalcescu, W~H Hsing, and S~Leibler.
\newblock {Resilient circadian oscillator revealed in individual
  cyanobacteria}.
\newblock {\em Nature}, 430(6995):81--85, 2004.

\bibitem{Zwicker2010}
David Zwicker, David~K Lubensky, and Pieter~Rein ten Wolde.
\newblock {Robust circadian clocks from coupled protein-modification and
  transcription-translation cycles. Supporting info}.
\newblock {\em Proceedings of the National Academy of Sciences of the United
  States of America}, 107(52):22540--5, December 2010.

\bibitem{Paijmans2015}
Joris Paijmans, Mark Bosman, Pieter~Rein Ten~Wolde, and David~K Lubensky.
\newblock {Discrete gene replication events drive coupling between the cell
  cycle and circadian clocks}.
\newblock {\em Submitted}, 2015.

\bibitem{Mugler:2010cq}
Andrew Mugler, Aleksandra Walczak, and Chris Wiggins.
\newblock {Information-Optimal Transcriptional Response to Oscillatory
  Driving}.
\newblock {\em Physical Review Letters}, 105(5):058101, July 2010.

\bibitem{Kageyama2003}
Hakuto Kageyama, Takao Kondo, and Hideo Iwasaki.
\newblock {Circadian formation of clock protein complexes by KaiA, KaiB, KaiC,
  and SasA in cyanobacteria.}
\newblock {\em J Biol Chem}, 278(4):2388--95, 2003.

\bibitem{Rust2007}
Michael~J Rust, Joseph~S Markson, William~S Lane, Daniel~S Fisher, and Erin~K
  O'Shea.
\newblock {Ordered phosphorylation governs oscillation of a three-protein
  circadian clock.}
\newblock {\em Science}, 318(5851):809--12, November 2007.

\bibitem{Nishiwaki2007}
Taeko Nishiwaki, Yoshinori Satomi, Yohko Kitayama, Kazuki Terauchi, Reiko
  Kiyohara, Toshifumi Takao, and Takao Kondo.
\newblock {A sequential program of dual phosphorylation of KaiC as a basis for
  circadian rhythm in cyanobacteria}.
\newblock {\em EMBO J}, 26:4029--4037, 2007.

\bibitem{Kitayama2008}
Yohko Kitayama, Taeko Nishiwaki, Kazuki Terauchi, and Takao Kondo.
\newblock {Dual KaiC-based oscillations constitute the circadian system of
  cyanobacteria}.
\newblock {\em Genes Dev}, 22:1513--1521, 2008.

\bibitem{Takai2006}
Naoki Takai, Masato Nakajima, Tokitaka Oyama, Ryotaku Kito, Chieko Sugita,
  Mamoru Sugita, Takao Kondo, and Hideo Iwasaki.
\newblock {A KaiC-associating SasA-RpaA two-component regulatory system as a
  major circadian timing mediator in cyanobacteria}.
\newblock {\em Proc Natl Acad Sci USA}, 103(32):12109--14, 2006.

\bibitem{Taniguchi:2007jx}
Y~Taniguchi, M~Katayama, R~Ito, N~Takai, T~Kondo, and T~Oyama.
\newblock {labA: a novel gene required for negative feedback regulation of the
  cyanobacterial circadian clock protein KaiC}.
\newblock {\em Gens Dev}, 21(1):60--70, January 2007.

\bibitem{Taniguchi2010}
Yasuhito Taniguchi, Naoki Takai, Mitsunori Katayama, Takao Kondo, and Tokitaka
  Oyama.
\newblock {Three major output pathways from the KaiABC-based oscillator
  cooperate to generate robust circadian kaiBC expression in cyanobacteria.}
\newblock {\em Proc Natl Acad Sci USA}, 107(7):3263--8, 2010.

\bibitem{Gutu2013}
Andrian Gutu and Erin~K O'Shea.
\newblock {Two antagonistic clock-regulated histidine kinases time the
  activation of circadian gene expression.}
\newblock {\em Mol Cell}, 50(2):288--294, March 2013.

\bibitem{Markson2013}
Joseph~S Markson, Joseph~R Piechura, Anna~M Puszynska, and Erin~K O'Shea.
\newblock {Circadian control of global gene expression by the cyanobacterial
  master regulator RpaA.}
\newblock {\em Cell}, 155(6):1396--408, 2013.

\bibitem{Tkacik2010}
Gasper Tkacik, Jason~S Prentice, Vijay Balasubramanian, and Elad Schneidman.
\newblock {Optimal population coding by noisy spiking neurons.}
\newblock {\em Proceedings of the National Academy of Sciences of the United
  States of America}, 107(32):14419--24, August 2010.

\bibitem{Walczak:2010cv}
Aleksandra~M Walczak, Ga{\v s}per Tka{\v c}ik, and William Bialek.
\newblock {Optimizing information flow in small genetic networks. II.
  Feed-forward interactions}.
\newblock {\em Physical Review E}, 81(4):041905, April 2010.

\bibitem{Dubuis2013}
Julien~O Dubuis, Gasper Tkacik, Eric~F Wieschaus, Thomas Gregor, and William
  Bialek.
\newblock {Positional information, in bits.}
\newblock {\em Proceedings of the National Academy of Sciences of the United
  States of America}, 110(41):16301--8, October 2013.

\bibitem{paulsson2003}
Johan Paulsson.
\newblock Summing up the noise in gene networks.
\newblock {\em Nature}, 427:415, 2004.

\bibitem{berg1977}
Howard~C. Berg and Edward~M. Purcell.
\newblock Physics of chemoreception.
\newblock {\em Biophysical Journal}, 20:193, 1977.

\bibitem{Ueda:2007uq}
Masahiro Ueda and Tatsuo Shibata.
\newblock Stochastic signal processing and transduction in chemotactic response
  of eukaryotic cells.
\newblock {\em Biophysical Journal}, 93(1):11--20, 2007.

\bibitem{bialek2005}
William Bialek and Sima Setayeshgar.
\newblock Physical limits to biochemical signaling.
\newblock {\em Proceedings of the National Academy of Sciences USA}, 102:10040,
  2005.

\bibitem{levinepre2007}
Kai Wang, Wouter-Jan Rappel, Rex Kerr, and Herbert Levine.
\newblock Quantifying noise levels in intercellular signals.
\newblock {\em Physical Review E}, 75:061905, 2007.

\bibitem{levineprl2008}
Wouter-Jan Rappel and Herbert Levine.
\newblock Receptor noise and directional sensing in eukaryotic chemotaxis.
\newblock {\em Physical Review Letters}, 100:228101, 2008.

\bibitem{wingreen2009}
Robert~G. Endres and Ned~S. Wingreen.
\newblock Maximum likelihood and the single receptor.
\newblock {\em Physical Review Letters}, 103:158101, 2009.

\bibitem{levineprl2010}
Bo~Hu, Wen Chen, Wouter-Jan Rappel, and Herbert Levine.
\newblock Physical limits on cellular sensing of spatial gradients.
\newblock {\em Physical Review Letters}, 105:048104, 2010.

\bibitem{mora2010}
Thierry Mora and Ned~S. Wingreen.
\newblock Limits of sensing temporal concentration changes by single cells.
\newblock {\em Physical Review Letters}, 104:248101, 2010.

\bibitem{Govern2012}
Christopher~C Govern and Pieter~Rein ten Wolde.
\newblock Fundamental limits on sensing chemical concentrations with linear
  biochemical networks.
\newblock {\em Physical Review Letters}, 109(21):218103, 2012.

\bibitem{Mehta2012}
Pankaj Mehta and David~J Schwab.
\newblock Energetic costs of cellular computation.
\newblock {\em Proceedings of the National Academy of Sciences USA},
  109(44):17978--17982, 2012.

\bibitem{Skoge:2011gi}
Monica Skoge, Yigal Meir, and Ned~S Wingreen.
\newblock {Dynamics of Cooperativity in Chemical Sensing among Cell-Surface
  Receptors}.
\newblock {\em Physical Review Letters}, 107(17):178101, October 2011.

\bibitem{Skoge:2013fq}
Monica Skoge, Sahin Naqvi, Yigal Meir, and Ned~S Wingreen.
\newblock Chemical sensing by nonequilibrium cooperative receptors.
\newblock {\em Physical Review Letters}, 110(24):248102, June 2013.

\bibitem{Kaizu:2014eb}
Kazunari Kaizu, Wiet de~Ronde, Joris Paijmans, Koichi Takahashi, Filipe
  Tostevin, and Pieter~Rein ten Wolde.
\newblock The berg-purcell limit revisited.
\newblock {\em Biophysical journal}, 106(4):976--985, 2014.

\bibitem{Govern:2014ef}
Christopher~C Govern and Pieter~Rein ten Wolde.
\newblock {Optimal resource allocation in cellular sensing systems}.
\newblock {\em Proceedings of the National Academy of Sciences of the United
  States of America}, 111(49):17486--17491, December 2014.

\bibitem{Govern:2014ez}
Christopher~C Govern and Pieter~Rein ten Wolde.
\newblock {Energy Dissipation and Noise Correlations in Biochemical Sensing}.
\newblock {\em Physical Review Letters}, 113(25):258102, December 2014.

\bibitem{Lang:2014ir}
Alex~H Lang, Charles~K Fisher, Thierry Mora, and Pankaj Mehta.
\newblock {Thermodynamics of Statistical Inference by Cells}.
\newblock {\em Physical Review Letters}, 113(14):148103, October 2014.

\bibitem{Ziv2007}
Etay Ziv, Ilya Nemenman, and Chris~H. WIggins.
\newblock {Optimal signal processing in small stochastic biochemical networks.}
\newblock {\em PloS one}, 2(10):e1077, January 2007.

\bibitem{Tostevin2009}
Filipe Tostevin and Pieter~Rein ten Wolde.
\newblock {Mutual Information between Input and Output Trajectories of
  Biochemical Networks}.
\newblock {\em Physical Review Letters}, 102(21):218101, May 2009.

\bibitem{Mehta2009}
Pankaj Mehta, Sidhartha Goyal, Tao Long, Bonnie~L. Bassler, and Ned~S.
  Wingreen.
\newblock {Information processing and signal integration in bacterial quorum
  sensing.}
\newblock {\em Molecular systems biology}, 5(325):325, January 2009.

\bibitem{Tkacik:2009ta}
Ga{\v s}per Tka{\v c}ik, Aleksandra~M Walczak, and William Bialek.
\newblock {Optimizing information flow in small genetic networks.}
\newblock {\em Physical Review E}, 80(3 Pt 1):031920, September 2009.

\bibitem{tostevin10}
Filipe Tostevin and Pieter~Rein {ten Wolde}.
\newblock Mutual information in time-varying biochemical systems.
\newblock {\em Phys Rev E Stat Nonlin Soft Matter Phys}, 81(6 Pt 1):061917, Jun
  2010.

\bibitem{DeRonde2010}
Wiet de~Ronde, Filipe Tostevin, and Pieter~Rein ten Wolde.
\newblock {Effect of feedback on the fidelity of information transmission of
  time-varying signals}.
\newblock {\em Physical Review E}, 82(3), September 2010.

\bibitem{DeRonde2011}
Wiet de~Ronde, Filipe Tostevin, and Pieter ten Wolde.
\newblock {Multiplexing Biochemical Signals}.
\newblock {\em Physical Review Letters}, 107(4):1--4, July 2011.

\bibitem{Cheong:2011jp}
R~Cheong, A~Rhee, C~J Wang, I~Nemenman, and A~Levchenko.
\newblock {Information Transduction Capacity of Noisy Biochemical Signaling
  Networks}.
\newblock {\em Science}, 334(6054):354--358, October 2011.

\bibitem{deRonde:2012fs}
W~de~Ronde, F~Tostevin, and P~ten Wolde.
\newblock {Feed-forward loops and diamond motifs lead to tunable transmission
  of information in the frequency domain}.
\newblock {\em Physical Review E}, 86(2):021913, August 2012.

\bibitem{Bowsher:2013jh}
Clive~G Bowsher, Margaritis Voliotis, and Peter~S Swain.
\newblock {The fidelity of dynamic signaling by noisy biomolecular networks.}
\newblock {\em PLoS Computational Biology}, 9(3):e1002965, 2013.

\bibitem{Selimkhanov:2014gd}
Jangir Selimkhanov, Brooks Taylor, Jason Yao, Anna Pilko, John Albeck,
  Alexander Hoffmann, Lev Tsimring, and Roy Wollman.
\newblock {Systems biology. Accurate information transmission through dynamic
  biochemical signaling networks.}
\newblock {\em Science}, 346(6215):1370--1373, December 2014.

\bibitem{DeRonde:2014fq}
Wiet de~Ronde and Pieter Rein~ten Wolde.
\newblock {Multiplexing oscillatory biochemical signals}.
\newblock {\em Physical Biology}, 11(2):026004, April 2014.

\bibitem{Sokolowski:2015km}
Thomas~R Sokolowski and Ga{\v s}per Tka{\v c}ik.
\newblock {Optimizing information flow in small genetic networks. IV. Spatial
  coupling}.
\newblock {\em Physical Review E}, 91(6):062710, June 2015.

\bibitem{Becker2015}
Nils~B Becker, Andrew Mugler, and Pieter~Rein ten Wolde.
\newblock {Optimal Prediction by Cellular Signaling Networks}.
\newblock {\em Physical Review Letters}, Accepted, 2015.

\bibitem{Tkacik2011}
Ga\v{s}per Tka\v{c}ik and Aleksandra~M Walczak.
\newblock {Information transmission in genetic regulatory networks: a review.}
\newblock {\em Journal of physics. Condensed matter : an Institute of Physics
  journal}, 23(15):153102, April 2011.

\bibitem{Shannon1948}
C.~E. Shannon.
\newblock {The mathematical theory of communication. 1963.}
\newblock {\em M.D. computing : computers in medical practice}, 14(4):306--17,
  1948.

\bibitem{Taniguchi:2010cb}
Y~Taniguchi, P~J Choi, G~W Li, H~Chen, M~Babu, J~Hearn, A~Emili, and X~S Xie.
\newblock {Quantifying E. coli Proteome and Transcriptome with Single-Molecule
  Sensitivity in Single Cells}.
\newblock {\em Science}, 329(5991):533--538, July 2010.

\bibitem{tanase-nicola06}
Sorin Tănase-Nicola, Patrick~B. Warren, and Pieter~Rein ten Wolde.
\newblock Signal detection, modularity, and the correlation between extrinsic
  and intrinsic noise in biochemical networks.
\newblock {\em Physical Review Letters}, 97(6):068102, August 2006.

\end{thebibliography}

\end{document}